\documentclass[%
 reprint,
superscriptaddress,
nobibnotes,
 amsmath,amssymb,
 aps,
prl
]{revtex4-2}

\usepackage{amsmath}
\usepackage{amssymb}
\usepackage{amsthm}
\usepackage{physics}
\usepackage{graphicx}
\usepackage{dcolumn}
\usepackage{bm}
\usepackage{xcolor}
\usepackage{hyperref}

\theoremstyle{definition}

\begin{document}

\preprint{APS/123-QED}

\title{Digital Coherent Twin-Field Quantum Key Distribution with Gaussian-Modulated Coherent States}

\author{Makoto Ishihara}
 \email{llmakomako.arg1076@keio.jp}
 \affiliation{%
 Department of Electronics and Electrical Engineering, Keio University, 3-14-1 Hiyoshi, Kohoku-ku, Yokohama 223-8522, Japan
}%
\author{Kazufumi Tanji}
 \affiliation{%
 Department of Electronics and Electrical Engineering, Keio University, 3-14-1 Hiyoshi, Kohoku-ku, Yokohama 223-8522, Japan
}%
\author{Wojciech Roga}%
 \affiliation{%
 Department of Electronics and Electrical Engineering, Keio University, 3-14-1 Hiyoshi, Kohoku-ku, Yokohama 223-8522, Japan
}%
\author{Jonas S. Neergaard-Nielsen}
\affiliation{%
 Center for Macroscopic Quantum States (bigQ), Department of Physics, Technical University of Denmark, 2800 Kongens Lyngby, Denmark
}%
\author{Ulrik L. Andersen}
\affiliation{%
 Center for Macroscopic Quantum States (bigQ), Department of Physics, Technical University of Denmark, 2800 Kongens Lyngby, Denmark
}%
\author{Masahiro Takeoka}%
 \email{takeoka@elec.keio.ac.jp}
\affiliation{%
 Department of Electronics and Electrical Engineering, Keio University, 3-14-1 Hiyoshi, Kohoku-ku, Yokohama 223-8522, Japan
}%
\affiliation{
National Institute of Information and Communications Technology (NICT), Koganei, Tokyo 184-8795, Japan.
}

\date{\today}

\begin{abstract}
Quantum key distribution (QKD) enables information-theoretically secure key sharing between two distant parties. Twin-field QKD (TF-QKD) can achieve long-distance secure communication thanks to single-photon interference. However, it requires active compensation for experimental imperfections such as phase, polarization, and frequency fluctuations with expensive devices and this makes its experimental implementation challenging. Here, we propose a long-distance continuous-variable QKD (CV-QKD) protocol which is inspired by TF-QKD. Our protocol can remove these requirements by using Gaussian-modulated coherent states and digital signal processing which is widely used in recent digital coherent optical communication and experimental implementations of CV-QKD. For security analysis, we use a numerical optimization approach with a photon-number cutoff assumption. We propose a new set of constraints for numerical optimization which can efficiently constrain a highly non-Gaussian quantum state shared by two legitimate parties in our protocol. We show that the key rate of our protocol exceeds the key rate of a direct-transmission CV-QKD protocol at long distances and it can overcome the repeaterless bound.

\end{abstract}

\maketitle


\section{\label{sec:introduction}INTRODUCTION}

Quantum key distribution (QKD) enables information-theoretically secure communication between distant parties, but its performance over optical fiber is fundamentally constrained by photon loss~\cite{Bennett2014,Xu2020,Pirandola2020}. For direct transmission through a pure-loss channel of transmissivity $\eta$, the secret-key rate cannot exceed the repeaterless bound~\cite{Takeoka2014,Pirandola2017,Wilde2017}, which decreases linearly with $\eta$ at high loss. Twin-field QKD (TF-QKD) fundamentally changes this scaling by interfering weak optical fields from Alice and Bob at an intermediate station~\cite{Lucamarini2018}. Because each field propagates only through a fraction of the total Alice--Bob link, TF-QKD can attain the characteristic $\sqrt{\eta}$ rate-loss scaling of the single-repeater bound~\cite{Pirandola2019}, allowing the repeaterless bound to be surpassed without quantum memories. This principle has stimulated extensive theoretical~\cite{Ma2018,Tamaki2018,Wang2018,Lin2018,Cui2019,Curty2019,Maeda2019,Lu2019,Curras-Lorenzo2021} and experimental~\cite{Minder2019,Wang2019,Liu2019,Zhong2019,Fang2020,Chen2020,Pittaluga2021,Liu2021,Chen2021,Clivati2022,Wang2022,Chen2022,Zhou2023,Liu2023,Li2023,Chen2024} developments.

TF-QKD has so far been developed predominantly within a discrete-variable framework. Continuous-variable (CV) QKD, by contrast, exploits coherent states with continuous modulation and offers a natural interface with coherent optical communication technology~\cite{Grosshans2002,Pirandola2020}. Gaussian modulation is particularly attractive experimentally because phase, frequency, and polarization transformations can be estimated using reference signals and incorporated into digital signal processing (DSP), as demonstrated in increasingly advanced CV-QKD systems~\cite{Qi2015,Soh2015,Eriksson2019,Eriksson2020,Matsuura2021,Jain2022,Chen2023,Hajomer2024,Hajomer2024Continuous,Jain2024Future,Hajomer2025High,Hajomer2025Experimental,Hajomer2025Coexistence,Hajomer2026Chip}. This raises a fundamental question: \emph{can continuous Gaussian modulation be combined with the twin-field mechanism while retaining its favorable rate-loss scaling?}

Here we show that it can. We introduce digital coherent twin-field QKD (DC-TF-QKD), in which Alice and Bob independently prepare weak Gaussian-modulated coherent states and transmit them to an untrusted central station, Charlie. Charlie interferes the incoming fields on a balanced beamsplitter and performs on-off photodetection; events in which exactly one detector clicks herald the correlations from which the secret key is extracted. Although the transmitted states are Gaussian, the heralding measurement is intrinsically non-Gaussian. In the symmetric configuration, the probability of a signal-induced successful event is governed by the transmissivity of a half link and therefore scales as $\sqrt{\eta}$, while the normalized state conditioned on such an event remains finite in the high-loss limit. Consequently, the secret-key rate inherits the characteristic TF scaling and can exceed the repeaterless bound.

The combination of Gaussian modulation and non-Gaussian heralding has two further consequences. First, the circular symmetry of the Gaussian modulation allows phase fluctuations, polarization rotations, and frequency offsets between the two optical fields to be estimated from pilot signals and incorporated into classical postprocessing. The protocol can therefore exploit DSP techniques originally developed for coherent optical communication~\cite{Kikuchi2016Fundamentals} and subsequently adapted to CV-QKD~\cite{Qi2015,Soh2015,Matsuura2021,Jain2022,Chen2023,Hajomer2024}, rather than requiring these reference-frame fluctuations to be continuously cancelled by active optical locking. Second, the state shared by Alice and Bob after Charlie's successful detection is strongly non-Gaussian, so Gaussian optimality~\cite{Navascues2006,GarciaPatron2006} and covariance-matrix characterization alone are insufficient for the security analysis.

To address this challenge, we employ the numerical security framework of Refs.~\cite{Coles2016,Winick2018} and introduce a set of experimentally accessible higher-order constraints that efficiently characterize the relevant non-Gaussian correlations. Using these constraints, together with a photon-number cutoff~\cite{Lin2019}, we obtain numerical key-rate bounds that closely approach those obtained when the heralded state is completely known. Our results establish that continuous Gaussian modulation and non-Gaussian single-click heralding can be combined to obtain twin-field rate-loss scaling, while simultaneously providing a route toward digital recovery of the optical reference frame. This connects two previously distinct approaches to optical QKD and opens a hybrid continuous-variable/twin-field architecture for long-distance quantum communication.

\section{PROTOCOL}
\begin{figure}[tbp]
\includegraphics[keepaspectratio, scale=0.38]{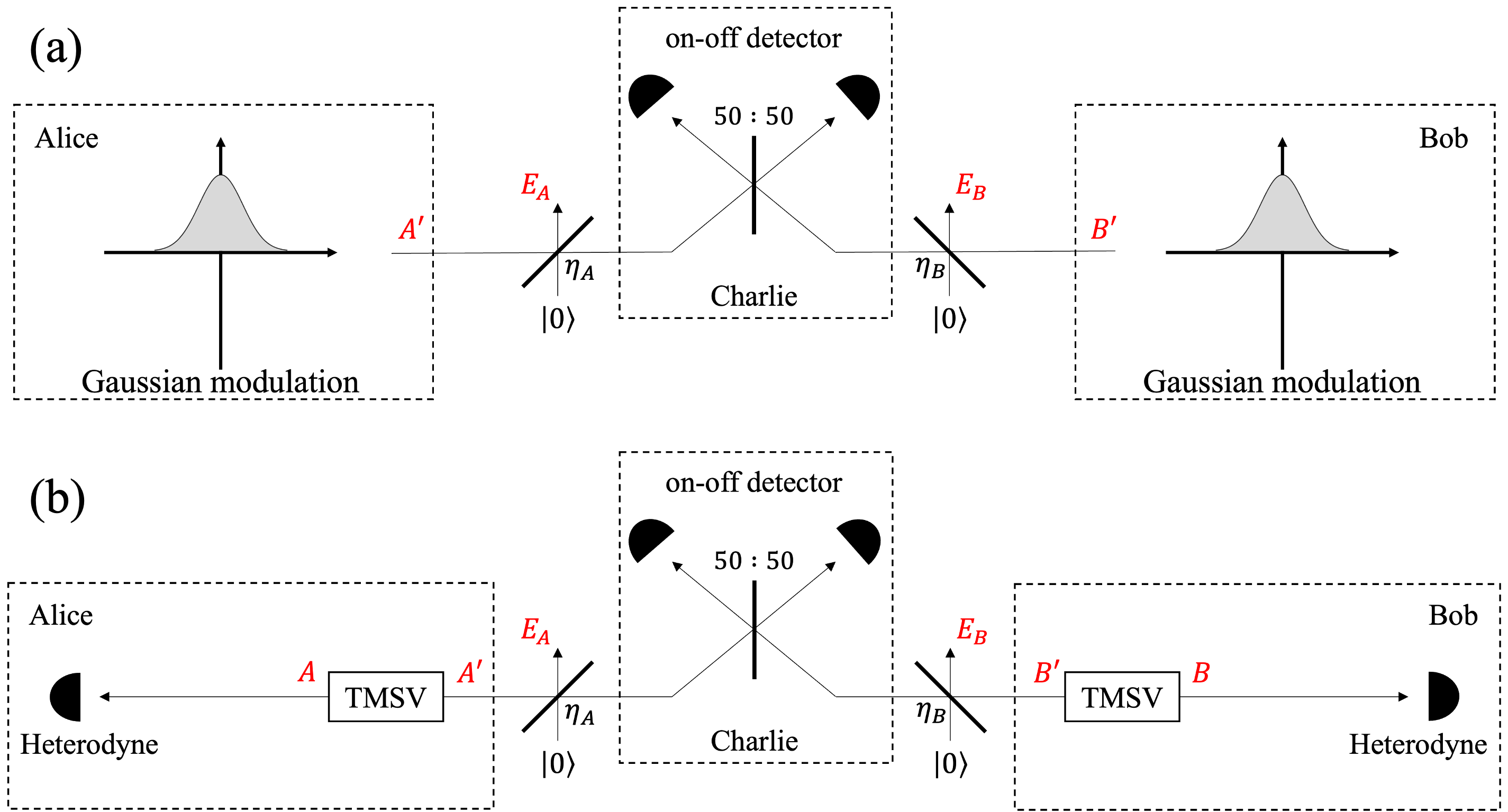}
\caption{\label{fig:CVQKDSchematic}
Schematic of DC-TF-QKD. (a) PM version: Alice and Bob prepare Gaussian-modulated coherent states and transmit them through pure-loss channels to Charlie, who interferes them on a 50:50 beamsplitter and performs on-off detection. Events in which exactly one detector clicks are retained. (b) Equivalent EB version: Alice and Bob prepare TMSV states and heterodyne their retained modes.}
\end{figure}

The basic DC-TF-QKD protocol is illustrated in Fig.~\ref{fig:CVQKDSchematic}.
We consider two equivalent variants: (a) a prepare-and-measure (PM) version and (b) an entanglement-based (EB) version.

In the PM version, shown in Fig.~\ref{fig:CVQKDSchematic}(a), Alice randomly prepares a coherent state from a Gaussian distribution of variance $V_A$ and sends it to an untrusted central node, Charlie, through a pure-loss channel with transmissivity $\eta_A$. Bob similarly prepares a coherent state from a Gaussian distribution of variance $V_B$ and sends it to Charlie through a channel with transmissivity $\eta_B$. Charlie interferes the two modes on a 50:50 beamsplitter and performs on-off detection at the two outputs. Events in which exactly one detector clicks are defined as successful, and Charlie publicly announces the click outcome. The detectors have efficiency $\eta_e^c$ and dark-count probability $p_\text{DC}$. Alice and Bob use a subset of the successful events for parameter estimation and, if a positive secret-key rate is certified, perform error correction and privacy amplification on the remaining data to obtain a secret key.

In the EB version, shown in Fig.~\ref{fig:CVQKDSchematic}(b), Alice and Bob each prepare a two-mode squeezed-vacuum (TMSV) state with mean photon numbers $\bar{n}_A$ and $\bar{n}_B$ per mode, respectively, and transmit one mode to Charlie. The mean photon numbers are related to the modulation variances by $\bar{n}_A=V_A/2$ and $\bar{n}_B=V_B/2$. Alice and Bob heterodyne their remaining modes and retain the outcomes whenever exactly one of Charlie's detectors clicks. They then perform parameter estimation, error correction, and privacy amplification as in the PM version. Assuming ideal heterodyne detection, the EB and PM descriptions are equivalent because heterodyne measurement of one mode of a TMSV prepares a Gaussian-modulated coherent state in the other mode.

Importantly, although Alice and Bob employ Gaussian state preparation, conditioning on Charlie's on-off detection produces a strongly non-Gaussian shared state. This heralding operation is central to both the twin-field rate-loss scaling and the security analysis discussed below.

\section{DIGITAL REFERENCE-FRAME COMPENSATION}
\begin{figure}[tbph]
\includegraphics[keepaspectratio, scale=0.28]{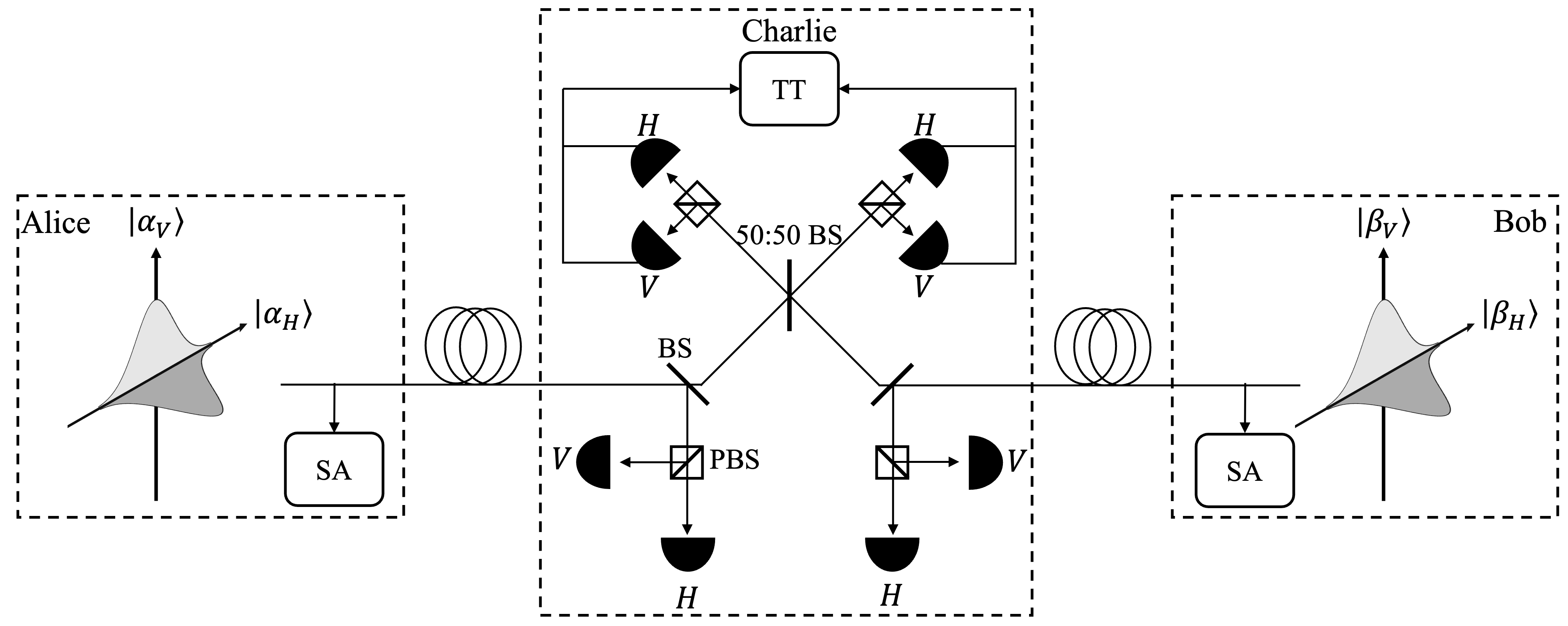}
\caption{\label{fig:CVQKDCompensation}
Digital compensation of phase, polarization, and frequency fluctuations in DC-TF-QKD. Pilot signals provide estimates of the relative phase and polarization transformations, while the frequency offset and detector time tag determine the frequency-induced phase rotation. These transformations are incorporated into Alice's and Bob's classical modulation data. BS, beamsplitter; PBS, polarizing beamsplitter; SA, spectrum analyzer; TT, time tagger.}
\end{figure}

In fiber transmission, optical phase and polarization fluctuate in time, while independent laser sources may also exhibit a frequency offset. In TF-QKD experiments, some or all of these degrees of freedom are typically actively stabilized~\cite{Pittaluga2021,Wang2022,Zhou2023,Liu2023,Chen2024}. In DC-TF-QKD, they can instead be monitored using pilot signals and compensated in classical postprocessing. This is enabled by the invariance of Gaussian modulation under phase rotations and, for equally modulated polarization modes, unitary polarization mixing. Figure~\ref{fig:CVQKDCompensation} illustrates the compensation scheme.

Alice and Bob transmit pilot pulses generated from the same laser sources as the quantum signals. At Charlie, the pilots are used to estimate the relative phase and polarization transformations, while the frequency difference can be monitored locally or estimated from the beat signal. The pilot signals may be time- or frequency-multiplexed with the quantum signals and separated before detection.

{\it Relative phase fluctuation}. Suppose Alice and Bob transmit coherent states $\ket{\alpha}$ and $\ket{\beta}$, which arrive at Charlie as $\ket{\alpha e^{i\phi_A}}$ and $\ket{\beta e^{i\phi_B}}$. Their interference depends on the relative phase $\Delta\phi=\phi_A-\phi_B$. By interfering pilot signals generated from the same optical sources, Charlie can estimate $\Delta\phi$. Alice and Bob then incorporate the corresponding phase rotation into their classical modulation data, e.g., by regarding Alice's transmitted amplitude as $\alpha e^{i\Delta\phi}$. Thus, the relative optical phase need not be actively locked during transmission.

{\it Polarization rotation}. Alice and Bob independently modulate two orthogonal polarization modes. For Alice, a polarization rotation transforms the coherent amplitudes according to
\begin{equation}
\bar{\alpha}_H=a\alpha_H+b\alpha_V,\qquad
\bar{\alpha}_V=-b^* \alpha_H+a^* \alpha_V,
\end{equation}
where $|a|^2+|b|^2=1$. By transmitting pilot pulses with known polarization and performing polarization-resolved measurements at Charlie, the coefficients $a$ and $b$ can be estimated and the corresponding transformation incorporated into Alice's classical modulation data. The same procedure applies to Bob.

{\it Frequency fluctuation}. Let $\omega_A$ and $\omega_B$ denote Alice's and Bob's optical frequencies, with $\Delta\omega=\omega_A-\omega_B$. Even when the optical spectra are displaced, interference can be recovered using sufficiently time-resolved detection~\cite{Zhao2014}. The frequency offset then appears as a detection-time-dependent phase $\Delta\omega t$. Hence, by estimating $\Delta\omega$ and recording the detection time $t$, Alice and Bob can account for this phase rotation in postprocessing. The relevant condition is that the phase varies negligibly within the detector time bin, $|\Delta\omega|\Delta t\ll1$. This regime is realistic for nanosecond-scale optical pulses, as commonly used in CV-QKD~\cite{Huang2015continuous,Ren2021demonstration,Wang2022continuous,Weerasinghe2023practical}, together with superconducting single-photon detectors having picosecond-scale timing resolution~\cite{Korzh2020Demonstration,Chang2021Detecting}. A formal derivation of the time-resolved interference is given in Supplemental Materical A~\cite{SM}.

These postprocessing operations are closely related to DSP techniques used in coherent optical communication and CV-QKD. They compensate reference-frame fluctuations but do not restore interference lost through irreversible mode mismatch; interference still requires the pulse envelopes to overlap in time and both fields to have a common spatial and polarization mode at the detector.

\section{KEY RATE}

The asymptotic secret-key rate of our protocol is given by the Devetak--Winter formula~\cite{Devetak2005},
\begin{equation}\label{KeyRateFormula1}
K_\text{key}=P_{\mathrm{succ}}
\left[
\beta I(\mathbf X;\mathbf Y)
-\max_{\rho_{AB}\in\mathbf S}\chi(\mathbf Y;E)
\right],
\end{equation}
where $P_{\mathrm{succ}}$ is the probability of a successful detection event at Charlie where, by symmetry, $P_\text{succ} = 2P_{10}$ with the probability that only the left detector clicks $P_{10}$, $I(\mathbf X;\mathbf Y)$ is the classical mutual information between Alice and Bob, $\chi(\mathbf{Y};E)$ is the Holevo information between Bob and Eve, $\beta$ is the reconciliation efficiency, and $\mathbf S$ is the set of states compatible with the observed data. For Gaussian CV-QKD, Eve's information can be bounded using Gaussian optimality~\cite{Navascues2006,GarciaPatron2006}. In DC-TF-QKD, however, conditioning on Charlie's successful on-off detection produces a strongly non-Gaussian state, for which first and second moments are insufficient.

We therefore employ the numerical security framework of Refs.~\cite{Coles2016,Winick2018}. Using the relative-entropy formulation of Ref.~\cite{Coles2012}, Eq.~\eqref{KeyRateFormula1} can be written as
\begin{equation}\label{KeyRate1}
K_\text{key}=P_{\mathrm{succ}}
\left\{
\min_{\rho_{AB}\in\mathbf S}
D\!\left(
\mathcal G(\rho_{AB})
\middle\|
\mathcal Z[\mathcal G(\rho_{AB})]
\right)
-\delta_{\mathrm{EC}}
\right\},
\end{equation}
where $\delta_{\mathrm{EC}}=H(\mathbf Y)-\beta I(\mathbf X;\mathbf Y)$. Here, $\mathcal G$ describes the key-map operation and $\mathcal Z$ dephases the corresponding key register. We discretize the heterodyne outcomes into $m$ angular regions; the resulting region operators, key map, and evaluation of the classical entropic quantities are given in Supplemental Material B~\cite{SM}.

To make the optimization finite dimensional, we adopt the photon-number cutoff assumption of Ref.~\cite{Lin2019}. Specifically, the local Hilbert spaces are truncated to $\{\ket{0},\ldots,\ket{N_c}\}$, such that
\begin{equation}
\rho=\Pi_{N_c}\rho\Pi_{N_c},
\qquad
\Pi_{N_c}=\sum_{n=0}^{N_c}\ketbra{n}.
\end{equation}
We use $N_c=5$ throughout; convergence with $N_c$ is examined in Supplemental Material B~\cite{SM}.

The feasible set is
\begin{equation}
\mathbf S=
\left\{
\rho_{AB}\ge0:
\mathrm{Tr}(\Gamma_i\rho_{AB})=\gamma_i,\ \forall i
\right\}.
\end{equation}
The central challenge is to choose observables that efficiently constrain the heralded non-Gaussian state. We use all bipartite combinations
\begin{equation}\label{eq:constraints}
\mathrm{Tr}
\!\left[
\rho_{AB}
(\hat O_A\otimes\hat O_B)
\right]
=
\langle\hat O_A\hat O_B\rangle ,
\end{equation}
with
\begin{equation}
\hat O\in
\{
\hat I,\hat q,\hat p,\hat q^2,\hat p^2,
\hat Q_{(2,1)},\hat P_{(2,1)},
\hat Q_{(3,2)},\hat P_{(3,2)}
\},
\end{equation}
where
\begin{align}
\hat Q_{(m,n)}
&=
(\hat a^\dagger)^m\hat a^n+
(\hat a^\dagger)^n\hat a^m,\\
\hat P_{(m,n)}
&=
i\!\left[
(\hat a^\dagger)^m\hat a^n-
(\hat a^\dagger)^n\hat a^m
\right].
\end{align}
The choice of these observables is motivated by the structure of the conditional state generated by the single-click measurement. For the honest implementation in the ideal one-photon heralding limit, Bob's state conditioned on Alice's outcome has, in the photon-number basis, nonzero matrix elements only on the diagonal and nearest off-diagonals, $\rho_{B|A}(i,j)=0$ for
$i\neq j,j\pm1$, and the same structure holds for $\rho_{A|B}$. The operators $\hat{Q}_{(m,m-1)}$ and $\hat{P}_{(m,m-1)}$ have support on these nearest off-diagonal elements and therefore directly constrain the photon-number coherences generated by the heralding measurement. The second-order operators $\hat{q}^2$ and $\hat{p}^2$, together with the identity, constrain the diagonal structure. Including several values of $m$ provides different photon-number weightings of the same coherences and significantly tightens the feasible set beyond what is possible from the covariance matrix alone. Importantly, all these expectation values remain directly accessible from the heterodyne statistics through the Husimi $Q$-function; details are provided in Supplemental Material B~\cite{SM}.

\section{RESULTS AND DISCUSSION}

We now evaluate the key rate of the elementary protocol in Fig.~\ref{fig:CVQKDSchematic}. The numerical results below do not include residual errors from the DSP operations discussed above. We consider the symmetric configuration $\eta_A=\eta_B=\sqrt{\eta}$, where $\eta=10^{-0.02L}$ and $L$ is the Alice--Bob distance in kilometers, and set $\bar n_A=\bar n_B=\bar n$. In addition to channel loss, we include the detector dark-count probability $p_{\mathrm{DC}}$, detection efficiency $\eta_e^c$, and reconciliation efficiency $\beta$. Unless otherwise stated, we use $m=16$ angular regions and optimize $\bar n$ numerically at each distance.

The origin of the twin-field scaling can be seen directly from the successful-event probability. Averaged over the Gaussian modulation, each mode arriving at Charlie is thermal. In the symmetric configuration, its mean photon number is $\mu=\bar{n}\sqrt{\eta}$. Neglecting dark counts for the scaling argument, the probability that an on-off detector with efficiency $\eta_e^c$ registers no click is
$q_0=(1+\eta_e^c\mu)^{-1}$. Since the balanced beamsplitter leaves two identical thermal inputs invariant, the probability that exactly one detector clicks is $P_{\mathrm{succ}}=2q_0(1-q_0) =2\eta_e^c\bar{n}\sqrt{\eta}+O(\eta)$. Moreover, in the high-loss limit, the unnormalized click-conditioned state and $P_{\mathrm{succ}}$ contain the same overall factor proportional to $\sqrt{\eta}$. After normalization, the heralded Alice--Bob state therefore approaches a finite nontrivial state. Consequently, provided that the secret fraction per successful event approaches a positive constant, the overall key rate $K_\text{key}=P_{\mathrm{succ}}r_{\mathrm{sec}}$ inherits the $\sqrt{\eta}$ scaling. The numerical optimization below confirms this behavior for the present protocol.

\begin{figure}[htbp]
 \centering
 \includegraphics[keepaspectratio,scale=0.4]{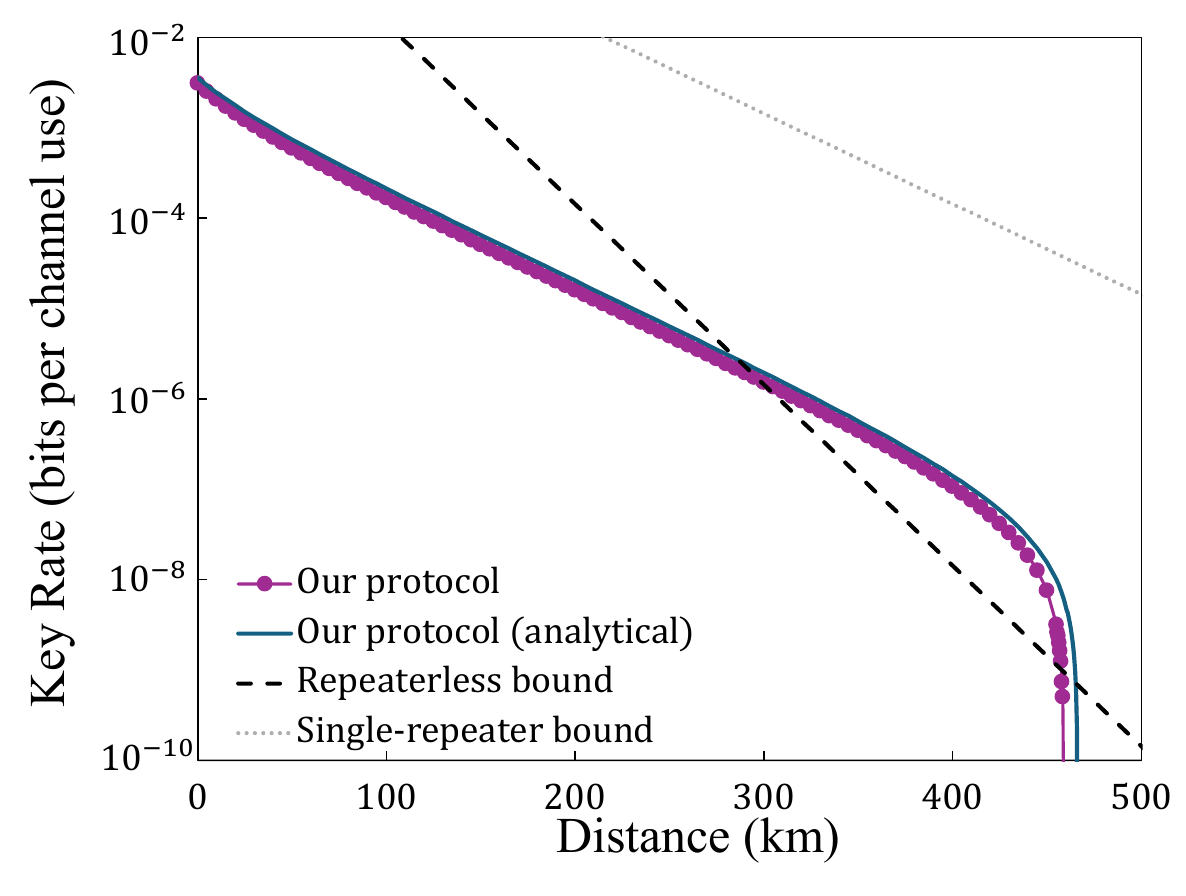}
 \caption{\label{fig:Mainplot}
Secret-key rate of DC-TF-QKD versus distance $L$. Purple circles show the numerical lower bound and the blue solid line the rate obtained when the conditional state is completely characterized, for $p_{\mathrm{DC}}=10^{-8}$, $\eta_e^c=0.7$, and $\beta=0.95$. The dashed and dotted black lines denote the repeaterless~\cite{Takeoka2014,Pirandola2017,Wilde2017} and single-repeater~\cite{Pirandola2019} bounds, respectively.}
\end{figure}

Figure~\ref{fig:Mainplot} shows the secret-key rate for $p_{\mathrm{DC}}=10^{-8}$, $\eta_e^c=0.7$, and $\beta=0.95$. The numerical lower bound obtained from the experimentally accessible constraints is compared with the rate obtained when Alice and Bob's conditional state is completely characterized. The two curves nearly coincide, demonstrating that the higher-order constraints introduced above tightly constrain the relevant non-Gaussian correlations.

Most importantly, the rate follows the characteristic twin-field scaling governed by the half-link transmissivity $\sqrt{\eta}$ and crosses the repeaterless bound~\cite{Takeoka2014,Pirandola2017,Wilde2017} at approximately $300$~km. Its loss exponent is therefore the same as that of the single-repeater bound~\cite{Pirandola2019}, in contrast to direct-transmission CV-QKD, whose rate is governed by the end-to-end transmissivity $\eta$~\cite{Lodewyck2007}. The numerical results thus confirm that continuous Gaussian modulation combined with non-Gaussian single-click heralding retains the fundamental rate-loss advantage of TF-QKD.

At sufficiently long distance, the performance is ultimately limited by detector dark counts: the signal-induced click probability decreases with $\sqrt{\eta}$, whereas the dark-count contribution is essentially distance independent. A reduction in detector efficiency mainly contributes additional effective loss and has a weaker influence on the maximum distance. Further dependence on $p_{\mathrm{DC}}$ and $\eta_e^c$ is presented in Supplemental Material C~\cite{SM}.

\section{CONCLUSION}

We have introduced DC-TF-QKD, which combines Gaussian-modulated coherent-state transmission with non-Gaussian single-click heralding at an untrusted central station. The protocol retains the characteristic twin-field rate-loss scaling and can surpass the repeaterless bound~\cite{Takeoka2014,Pirandola2017,Wilde2017}, demonstrating that the TF advantage is not intrinsically tied to discrete-variable encoding. Despite the strongly non-Gaussian heralded state, the proposed higher-order constraints yield numerical secret-key bounds that closely approach those obtained when the conditional state is completely known.

Gaussian modulation further enables phase, polarization, and frequency transformations to be monitored using pilot signals and incorporated into classical postprocessing, providing a route toward TF-QKD with digital reference-frame recovery. The present security analysis employs a photon-number cutoff; removing this assumption and extending the analysis to finite-size security~\cite{Leverrier2017,Matsuura2021,Kanitschar2023} are important directions for future work. More generally, developing efficient experimentally accessible constraints for non-Gaussian states may also be useful beyond QKD, where such states play a central role in continuous-variable quantum information processing~\cite{Walschaers2021}. Our results establish a hybrid continuous-variable/twin-field architecture combining coherent-state transmitters and digital signal processing with the favorable rate-loss scaling of non-Gaussian heralding.

\textit{Note added.}---While preparing this manuscript, we became aware of a related preprint by Pirandola~\cite{Pirandola2026CV}, which also combines Gaussian-modulated coherent states with single-photon interference. That work uses photon-number-resolving detection, whereas the present work uses on-off detection and phase, polarization, and frequency fluctuation compensation with digital signal processing and numerical key rates based on higher-order heterodyne moments.

\begin{acknowledgments}
\textit{Acknowledgments}---This work was supported by JST Moonshot R\&D, Grant No. JPMJMS226C and Grant No. JPMJMS256E, JST CRONOS, Grant No. JPMJCS24N6, JST COI-NEXT, Grant No. JPMJPF2221, and JST ASPIRE, Grant No. JPMJAP2427. We acknowledge support from the European Union’s Horizon Europe research and innovation programme under the project “Quantum Secure Networks Partnership” (QSNP, Grant Agreement No. 101114043), the Danish National Research Foundation through the Center for Macroscopic Quantum States (bigQ, DNRF142), the Innovation Fund Denmark through the CyberQ project (Grant Agreement No. 3200-00035B) and the AccessQKD project (Grant Agreement No. 43560-00004B). M.I. and K.T. were supported by JST SPRING, Grant No. JPMJSP2123.
\end{acknowledgments}

\bibliography{CVQKD}


\clearpage
\onecolumngrid

\begin{center}
\textbf{\large Supplemental Material}
\end{center}

\setcounter{equation}{0}
\renewcommand{\theequation}{S\arabic{equation}}

\section{A. TIME-RESOLVED INTERFERENCE OF TWO PULSES}\label{app:time-resolved}
In this section, we consider the interference of two coherent pulses with different central frequencies and show that even without spectrum overlap, they can effectively interfere by postprocessing compensation if the time resolution of the detector is much shorter than the pulse width.  

We consider a simple situation where two coherent states are interfered by a 50:50 beamsplitter and then the output is detected by on-off photon detectors. 
We start with the ideal case, i.e. the two coherent pulses, $|\alpha\rangle$ and $|\beta\rangle$, are exactly in the same mode. 
Since our DC-TF-QKD protocol uses the single-photon interference, let us see the event that only one of the detectors is clicked. 
The state after the beamsplitter turns out to be $|(\alpha+\beta)/\sqrt{2}\rangle|(\alpha-\beta)/\sqrt{2}\rangle$ and then the probability of detecting photons in the first mode and not in the second mode is given by 
\begin{equation}\label{ideal single-photon interference}
P_{10} = e^{-|\alpha|^2-|\beta|^2} \left\{ \exp \left( \frac{1}{2} |\alpha + \beta|^2 \right) -1 \right\} .
\end{equation}

The question asked here is the following. Consider the time and frequency distribution of these coherent pulses. 
Suppose the center frequencies of two pulses are not identical, e.g. due to the laser source fluctuation, but the users can monitor these frequencies. Meanwhile, we can use on-off detectors with very good time resolution, can tag the detection time, and separately monitor the frequency difference of two pulses. Then is it possible to cancel the frequency difference by postprocessing to get the desired interference? 
To consider this, define a creation operator in spectral mode of $f(\omega)$ as 
\begin{equation}\label{specral a dagger}
A_f^\dagger = \int d\omega \, f^*(\omega) a^\dagger (\omega) .
\end{equation}
Then the coherent state $|\alpha\rangle$ distributed in the spectrum of $f(t)$ is described as 
\begin{equation}\label{temporal coherent state alpha}
|\alpha : f\rangle = e^{-|\alpha|^2/2} \sum_{n=0}^\infty \frac{\alpha^{n}}{n!} A_f^{\dagger \, n} |0\rangle . 
\end{equation}
Similarly, $|\beta\rangle$ distributed in $g(\omega)$ is $|\beta : g\rangle$. Note that $\int d\omega \, |f(\omega)|^2 = \int d\omega \, |g(\omega)|^2 = 1$. 
Since we assume two laser sources have the same spectrum with different frequency, $g(\omega)=f(\omega + \delta\omega)$ holds.  
By using this, we can treat the single-photon interference in time and frequency domain. 
The beamsplitter mixes these coherent states as 
\begin{equation}\label{BS time resolved}
    |\alpha : f\rangle |\beta : g\rangle
    \to |\gamma_+ : u_+\rangle |\gamma_- : u_- \rangle ,
\end{equation}
where 
\begin{equation}\label{gamma_pm}
   \gamma_\pm = \sqrt{\frac{|\alpha|^2+|\beta|^2\pm 2 {\rm Re} \alpha\beta^* \langle f,g \rangle}{2}}, 
\end{equation}
and 
\begin{equation}\label{u_pm}
   u^*_\pm(\omega) = \frac{\pm \alpha f^*(\omega) + \beta g^*(\omega)}{\sqrt{|\alpha|^2+|\beta|^2\pm 2 {\rm Re} \alpha\beta^* \langle f,g \rangle}} .  
\end{equation}
To simplify the analysis, we consider on-off detection within a time interval $T = [t_0, t_0+\Delta t]$. The corresponding POVM elements are given by,
\begin{equation}
    \begin{split}
        \Pi_\text{off}^{(T)} = \ketbra{0}{0}_T \otimes I_{\bar{T}},\quad \Pi_\text{on}^{(T)} &= I - \Pi_\text{off}^{(T)},
    \end{split}
\end{equation}
where $\ket{0}_T$ denotes the vacuum state of all temporal modes within $T$ and $\bar{T}$ denotes the time region outside $T$. The photon-number densities at the outputs of the 50:50 beam splitter are
\begin{equation}
    \lambda_\pm (t) = \frac{1}{2}|\alpha^* f(t) \pm \beta^* g(t)|^2.
\end{equation}
Then, the mean photon numbers in the time interval $T$ become
\begin{equation}
    \mu_\pm (T) = \int_T dt ~ \lambda_\pm (t).
\end{equation}
Therefore, the probability that one detector clicks and the other one does not in $T$ is
\begin{equation}
    \begin{split}
        P_{10}(T) &= (1-e^{-\mu_+ (T)}) e^{-\mu_- (T)}\\
        &= e^{-(\mu_+(T) + \mu_-(T))} (e^{\mu_+ (T)} - 1).
    \end{split}
\end{equation}
Here, we have
\begin{equation}
    \begin{split}
        \mu_+ (T) + \mu_-(T) &= \int_T dt ~ (\lambda_+ (t) + \lambda_- (t))\\
        &= \int_T dt ~ (|\alpha|^2 |f(t)|^2 + |\beta|^2 |g(t)|^2)\\
        &= (|\alpha|^2 + |\beta|^2) \int_T dt ~ |f(t)|^2,
    \end{split}
\end{equation}
where we use $g(t) = \mathcal{F}^{-1} [f(\omega + \delta \omega)] = f(t) e^{- i \delta \omega t}$ for the last equality. Hence
\begin{equation}\label{eq:P10T}
    \begin{split}
        P_{10}(T) &= e^{- (|\alpha|^2 + |\beta|^2)F_T}  [\exp(\frac{1}{2} \int_T dt ~ |\alpha^* + \beta^* e^{- i \delta\omega t}|^2 |f(t)|^2)-1],
    \end{split}
\end{equation}
where
\begin{equation}
    F_T = \int_T dt ~ |f(t)|^2.
\end{equation}
Here, define
\begin{equation}
    J_T = \int_T dt ~ |\alpha^* + \beta^* e^{-i \delta \omega t}|^2 |f(t)|^2.
\end{equation}
Since 
\begin{equation}
    |\alpha^*+\beta^* e^{-i \delta \omega t}|^2 = |\alpha|^2 + |\beta|^2 + 2 \text{Re} [\alpha^* \beta e^{i \delta \omega t}],
\end{equation}
we have
\begin{equation}
    J_T = (|\alpha|^2 + |\beta|^2)F_T + 2 \text{Re}[\alpha^* \beta \int_T dt ~ e^{i \delta \omega t} |f(t)|^2].
\end{equation}
When the time resolution of the detector is sufficiently high such that $|\delta \omega| \Delta t \ll 1$, the following holds for $t \in T$:
\begin{equation}
    |\delta \omega |  (t - t_0) \leq |\delta \omega| \Delta t \ll 1,
\end{equation}
and consequently
\begin{equation}
    e^{i \delta \omega t} = e^{i \delta \omega t_0} e^{i \delta \omega (t- t_0)} \approx e^{i \delta \omega t_0}.
\end{equation}
Therefore,
\begin{equation}
    \begin{split}
        J_T &\approx (|\alpha|^2 + |\beta|^2)F_T + 2 \text{Re}[\alpha^* \beta e^{i \delta \omega t_0} F_T]\\
        &= F_T |\alpha + \beta e^{i \delta \omega t_0}|^2.
    \end{split}
\end{equation}
Finally, $P_{10} (T)$ can be approximated as
\begin{equation}
    P_{10}(T) \approx e^{-(|\alpha|^2 + |\beta|^2)F_T} [\exp(\frac{F_T}{2} |\alpha + \beta e^{i\delta \omega t_0}|^2)-1].
\end{equation}
This has the same form as Eq.~(\ref{ideal single-photon interference}), except that the coherent amplitudes are multiplied by $\sqrt{F_T}$ and the relative phase is shifted by $\delta \omega t_0$. It implies that, in the DC-TF-QKD protocol, by monitoring $\delta \omega$ and the detection time interval $T = [t_0, t_0 + \Delta t]$, one can account for the frequency difference by regarding Bob's coherent amplitude as $\beta e^{i \delta \omega t_0}$ instead of $\beta$. This phase correction can be incorporated into classical postprocessing. More detailed feasibility analysis is left for future work. Finally, we consider the limit where the time interval $T$ is sufficiently long to cover the entire pulse. In this limit, $F_T \rightarrow 1$ and Eq.~(\ref{eq:P10T}) becomes
\begin{equation}
    \begin{split}
        P_{10} &= e^{- (|\alpha|^2 + |\beta|^2)}  [\exp(\frac{1}{2} \int_{-\infty}^\infty dt ~ |\alpha^* + \beta^* e^{- i \delta\omega t}|^2 |f(t)|^2)-1].
    \end{split}
\end{equation}
Here, when $\delta \omega = 0$, the above expression exactly coincides with Eq.~(\ref{ideal single-photon interference}).


\section{B. DETAILS OF THE KEY RATE CALCULATION}\label{sec:KeyrateDetails}
We describe details about key rate calculation of our protocol. First, we reproduce Eq.~(\ref{KeyRate1}) in the main text, 
\begin{equation}\label{KeyRate}
    K_\text{key} = P_\text{succ} \{\min_{\rho_{AB} \in \mathbf{S}} D(\mathcal{G}(\rho_{AB})||\mathcal{Z}[\mathcal{G}(\rho_{AB})])- \delta_\text{EC} \}.
\end{equation}
$\mathcal{G} (\rho) = K \rho K^\dagger $ is a completely positive and trace-nonincreasing map describing the key-map operation, where
\begin{equation}
    K = \sum_{y=0}^{m-1} \ket{y}_R \otimes I_A \otimes (\sqrt{R_{y}})_B
\end{equation}
is a Kraus operator. Here, $R$ is a register system and $R_{j}$ are region operators defined as
\begin{equation}
    R_j = \frac{1}{\pi} \int_{0}^{\infty} \int_{(2j-1)\pi/m}^{(2j+1)\pi/m} \gamma \ketbra{\gamma e^{i\theta}} d\theta d\gamma,
\end{equation}
where $j \in \left\{ 0, 1, \ldots, m-1\right\}$ and $m$ is the number of region operators. $\mathcal{Z}$ is a pinching quantum channel that dephases the register $R$ to read out the result of a key map:
\begin{equation}
    \mathcal{Z}(\rho) = \sum_{j=0}^{m-1} Z_j \rho Z_j,
\end{equation}
where
\begin{equation}
    Z_j = \ketbra{j}_R \otimes I_A \otimes I_B
\end{equation}
is a projector. We can calculate the classical entropic functions by using the probability distribution $p(x=j,y=k) = \text{Tr}\left[ (R_j)_A (R_k)_B \rho_{AB}\right] $.

\begin{figure}[tbp]
 \centering
 \includegraphics[keepaspectratio, scale=0.4]{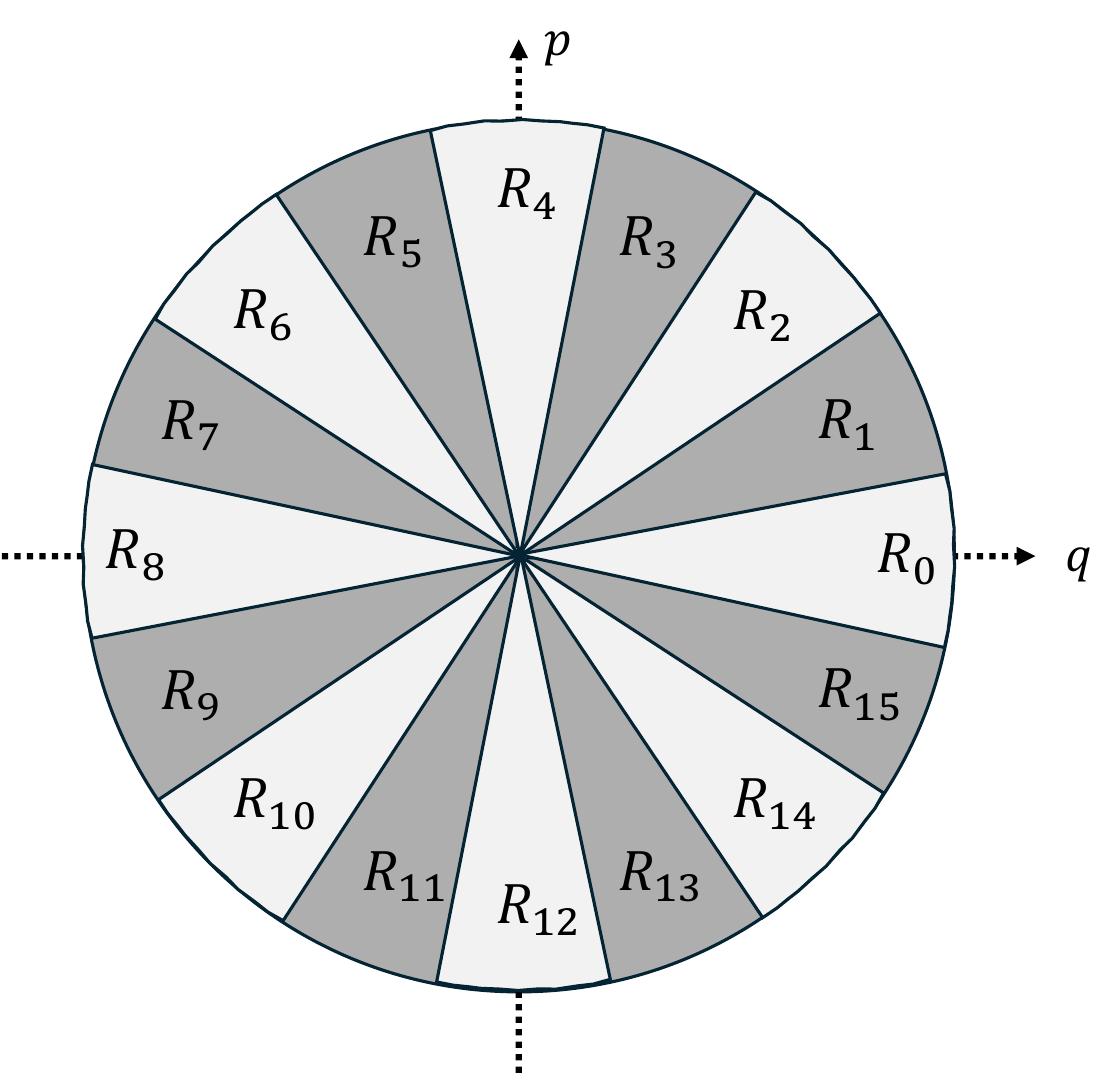}
 \caption{Region Operator $R_j$ where $m = 16$.}
 \label{fig:RegionOperator}
\end{figure}

\subsection{1. Photon-number cutoff}

In Eq.~(\ref{KeyRate}), we perform the optimization over all $\rho_{AB}$ in the feasible set $\mathbf{S}$ for estimating key rate. However, because $\rho_{AB}$ acts on an infinite-dimensional Hilbert space, we cannot perform this optimization numerically. Therefore, we make $\rho_{AB}$ finite dimensional. In our protocol, Alice's Hilbert space $\mathcal{H}_A$ and Bob's Hilbert space $\mathcal{H}_B$ which are spanned by the photon-number basis ($\ket{n}, n \in \{ 0, 1, \ldots, \infty \}$) are infinite dimensional. Then, we make $\mathcal{H}_A$ and $\mathcal{H}_B$ finite dimensional by using the photon-number cutoff assumption~\cite{Lin2019}, that is, we assume that $\mathcal{H}_A$ and $\mathcal{H}_B$ are spanned by a basis $\{ \ket{0}, \ket{1}, \ldots, \ket{N_c} \}$, where $N_c$ is a cutoff photon number. If we define $\Pi_{N_c} = \sum_{n=0}^{N_c} \ketbra{n}$, a quantum state $\rho$ under this assumption is $\rho = \Pi_{N_c} \rho \Pi_{N_c}$. 

The annihilation and creation operators in the photon-number basis up to cutoff photon number $N_c$ are, respectively,
\begin{equation}\label{annihilationandcreation}
\begin{split}
    &\Pi_{N_c} \hat{a} \Pi_{N_c}^\dagger = \left[  
    \begin{matrix}
        0 & 1 & 0 & \cdots & 0\\
        0 & 0 & \sqrt{2} & \cdots & 0\\
        \vdots & & & \ddots & \vdots\\
        0 & 0 & 0 & \cdots & \sqrt{N_c}\\
        0 & 0 & 0 & \cdots & 0
    \end{matrix}
    \right],\quad \Pi_{N_c} \hat{a}^\dagger \Pi_{N_c}^\dagger = \left[  
    \begin{matrix}
        0 & 0 &  \cdots & 0 & 0\\
        1 & 0 &  & 0 & 0\\
        0 & \sqrt{2} & & 0 & 0\\
        \vdots & \vdots & \ddots & \vdots & \vdots\\
        0 & 0 & \cdots & \sqrt{N_c} & 0
    \end{matrix}
    \right].
\end{split}
\end{equation}
We can write the region operators $R_j$ in this basis, shown in Fig.~\ref{fig:RegionOperator}. The overlap between a coherent state and a photon-number state is as follows:
\begin{equation}
    \braket{re^{i\theta}}{n} = e^{-(r^2/2)} r^n e^{-in\theta}/\sqrt{n!}.
\end{equation}
Therefore, the matrix elements of $R_j$ in the photon-number basis are, for $k \neq l$,
\begin{equation}
    \begin{split}
        \mel{k}{R_j}{l} &= \frac{1}{\pi} \int_0^\infty \int_{(2j-1)\pi/m}^{(2j+1)\pi/m} \braket{k}{re^{i\theta}} \braket{re^{i\theta}}{l} r d\theta dr\\
        &= \frac{i}{2\pi(k-l)\sqrt{k!l!}}\Gamma \left(\frac{k+l+2}{2} \right)\left(e^{i(k-l)(2j-1)\pi/m} - e^{i(k-l)(2j+1)\pi/m}  \right),
    \end{split}
\end{equation}
where $\Gamma$ is the gamma function. In particular, the diagonal elements are
\begin{equation}
    \begin{split}
        \mel{k}{R_j}{k} &= \frac{1}{\pi k!} \int_0^\infty r^{2k+1} e^{-r^2} dr \int_{(2j-1)\pi/m}^{(2j+1)\pi/m} d\theta\\
        &= \frac{1}{m},
    \end{split}
\end{equation}

\begin{figure}[tbp]
 \centering
 \includegraphics[keepaspectratio, scale=0.37]{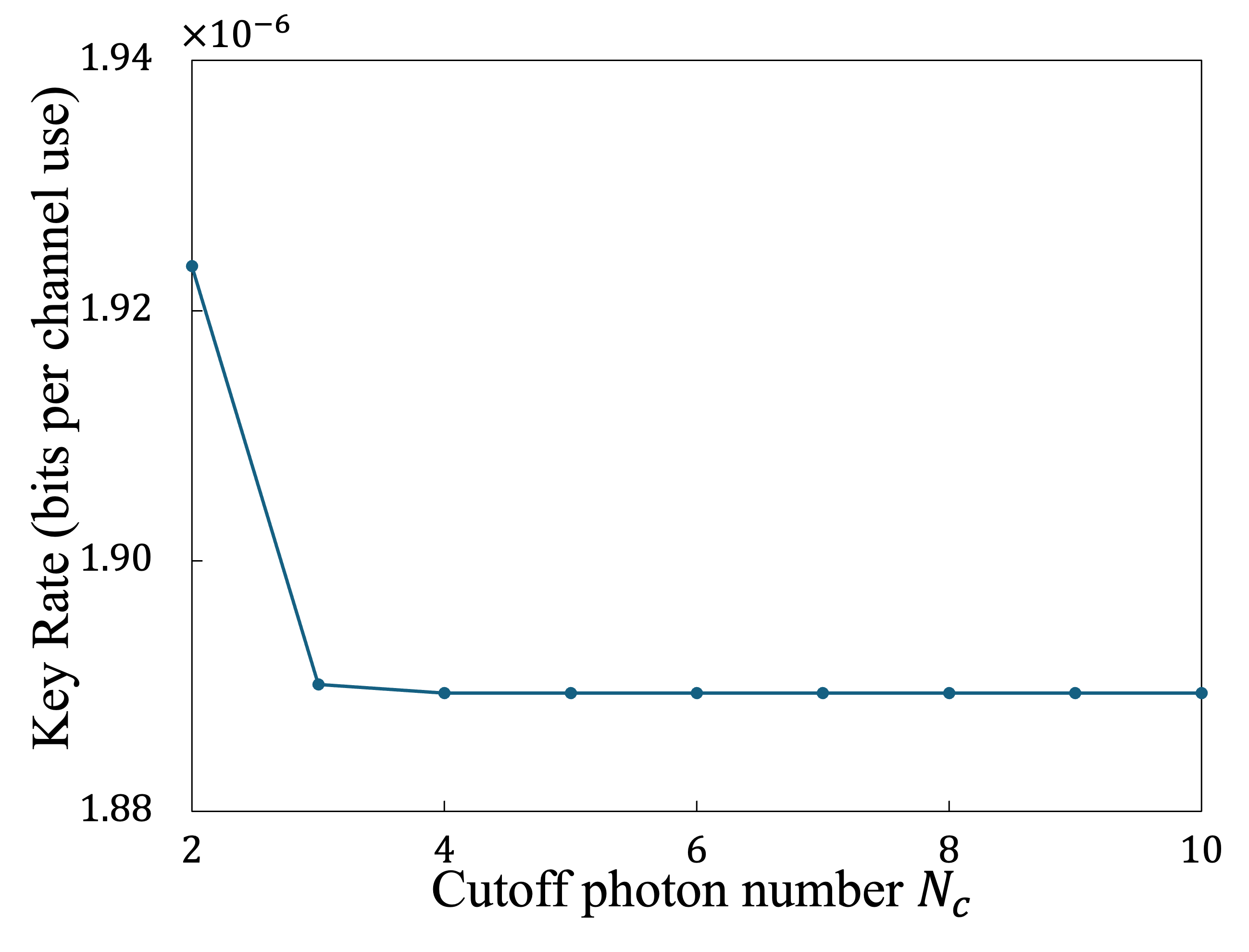}
 \caption{Key rate versus cutoff photon number $N_c$. We fix $L = 300$ km and $\bar{n} = 0.012$. Other parameters are $p_\text{DC} = 10^{-8}, \eta_e^c = 0.7$, and $\beta = 0.95$.}
 \label{fig:Cutoff}
\end{figure}

In this paper, we use the cutoff photon number $N_c = 5$. To justify this choice, we plot analytically calculated key rates of our protocol for different cutoff $N_c$ in Fig.~\ref{fig:Cutoff}. While the key rate decreases as the cutoff photon number increases, it saturates when $N_c$ is larger than 3. Note that this convergence check does not remove the photon-number cutoff assumption used in the security analysis.

\subsection{2. Constraints}
We describe the feasible set $\mathbf{S}$ in the optimization problem in Eq.~(\ref{KeyRate}). $\mathbf{S}$ is expressed as 
\begin{equation}
    \mathbf{S} = \{ \rho_{AB} \geq 0 | {\text{Tr}}[\Gamma_i \rho_{AB}] = \gamma_i, \,  \forall i  \}.
\end{equation}
We choose the following constraints:
\begin{equation}
    \begin{split}
        \rho_{AB} &\geq 0\\
        {\text{Tr}} [\rho_{AB} (\hat{O}_A \otimes \hat{O}_B)] &=\ev{\hat{O}_A \hat{O}_B},
    \end{split}
\end{equation}
where $\hat{O} \in \{ \hat{I}, \hat{q}, \hat{p}, \hat{q}^2, \hat{p}^2, \hat{Q}_{(2,1)}, \hat{P}_{(2,1)}, \hat{Q}_{(3,2)}, \hat{P}_{(3,2)} \}$, and we use all combinations of $\hat{O}_A$ and $\hat{O}_B$ as the constraints. 

Here, we explain why we choose these operators. Let $\rho(i, j)$ denote the $(i, j)$ element of a density matrix $\rho$. In addition, let $\rho_{B|A}$ denote Bob's quantum state in the photon-number basis conditioned on Alice's heterodyne measurement outcome  in an honest implementation in the ideal one-photon heralding limit. Then, for the elements of $\rho_{B|A}$, the following holds.
\begin{equation}
    \rho_{B|A} (i, j) = 0 \quad \text{if} \, i \neq j,  j \pm 1.
\end{equation}
That is, the density matrix $\rho_{B|A}$ has the form in the photon-number basis as follows
\begin{equation}
    \rho_{B|A} = \left[  
    \begin{matrix}
        a_0 & b_0 & 0 & \cdots & 0\\
        b_0^* & a_1 & b_1 & \ddots & \vdots\\
        0 & b_1^* & \ddots & \ddots & 0\\
        \vdots & \ddots & \ddots & \ddots & b_{N_c-1}\\
        0 & \cdots & 0 & b_{N_c-1}^* & a_{N_c}
    \end{matrix}
    \right],
\end{equation}
where $a_0, a_1, \ldots  a_{N_c}$ are diagonal elements and $b_0, b_1, \ldots, b_{N_c-1}$ are nearest off-diagonal elements. From the symmetry of the protocol, Alice's quantum state conditioned on Bob's measurement outcome $\rho_{A|B}$ has the same structure. Therefore, we should choose operators which can efficiently constrain these diagonal and nearest off-diagonal elements. We consider the operators $\hat{Q}_{(m,n)}$ and $\hat{P}_{(m,n)}$ expressed by the photon-number basis. Let $\hat{Q}_{(m,n)}(i, j)$ and $\hat{P}_{(m,n)}(i,j)$ denote $(i, j)$ elements of $\hat{Q}_{(m,n)}$ and $\hat{P}_{(m,n)}$, respectively. Then, the following holds.
\begin{equation}
    \hat{Q}_{(m,m-1)}(i, j) = \hat{P}_{(m, m-1)} (i, j) = 0 \quad \text{if} \, i \neq j \pm 1.
\end{equation}
This means that these operators have only elements next to diagonal elements. Therefore, by using these operators, we can constrain the nearest off-diagonal elements $b_1 \ldots, b_{N_c-1}$ of the quantum states $\rho_{B|A}$ and $\rho_{A|B}$. Next, let $\hat{q}^2 (i, j)$ and $\hat{p}^2 (i, j)$ denote $(i, j)$ elements of the operators $\hat{q}^2$ and $\hat{p}^2$ in the photon-number basis, respectively. Then,
\begin{equation}
    \hat{q}^2(i, j) = \hat{p}^2 (i, j) = 0 \quad \text{if} \, i \neq j, j \pm 2.
\end{equation}
Thus, these operators are useful for estimating the diagonal elements $a_1, \ldots, a_{N_c}$. Therefore, the set of the operators $\{ \hat{I}, \hat{q}, \hat{p}, \hat{q}^2, \hat{p}^2, \hat{Q}_{(2,1)}, \hat{P}_{(2,1)}, \hat{Q}_{(3,2)}, \hat{P}_{(3,2)} \}$ can constrain a quantum state shared between Alice and Bob in our protocol.

\subsection{3. Calculation of expectation values}
The operators in the set $\{ \hat{I}, \hat{q}, \hat{p}, \hat{q}^2, \hat{p}^2, \hat{Q}_{(2,1)}, \hat{P}_{(2,1)}, \hat{Q}_{(3,2)}, \hat{P}_{(3,2)} \}$ can be expressed by using the annihilation and creation operators in the following way
\begin{equation}
    \hat{f} (\hat{a}, \hat{a}^\dagger) = \sum_{n,m} d_{nm} (\hat{a})^n (\hat{a}^\dagger)^m,
\end{equation}
where $d_{nm}$ are coefficients. For example, the quadrature operators $\hat{q}$ and $\hat{p}$ are expressed as
\begin{equation}
    \hat{q} = \frac{1}{\sqrt{2}} (\hat{a} + \hat{a}^\dagger), \quad
    \hat{p} = \frac{i}{\sqrt{2}} (- \hat{a} + \hat{a}^\dagger).
\end{equation}
Also, the operators $\hat{Q}_{(m,n)}$ and $\hat{P}_{(m,n)}$ are expressed as 
\begin{equation}
    \hat{Q}_{(m,n)} = (\hat{a}^\dagger)^m (\hat{a})^n + (\hat{a}^\dagger)^n (\hat{a})^m, \quad \hat{P}_{(m,n)} = i\{ (\hat{a}^\dagger)^m (\hat{a})^n - (\hat{a}^\dagger)^n (\hat{a})^m \}.
\end{equation}
By using the commutation relation between the annihilation and the creation operators, we can change the order of these operators in $\hat{Q}_{(m,n)}$ and $\hat{P}_{(m,n)}$.

Here, if Alice and Bob perform heterodyne measurements on a density operator $\rho$, they obtain the Husimi $Q$-function
\begin{equation}
    Q (\alpha, \beta) = \mel{\alpha \beta}{\rho}{\alpha \beta}/\pi^2,
\end{equation}
where $\ket{\alpha}$ and $\ket{\beta}$ are Alice's and Bob's measurement outcomes, respectively. Then, we can calculate expectation values of the operator used for estimating a quantum state between Alice and Bob by using the following equation \cite{Scully1997}:
\begin{equation}
    \ev{\hat{f} (\hat{a}_A, \hat{a}_A^\dagger) \hat{f} (\hat{a}_B, \hat{a}_B^\dagger)} = \int \int \int \int Q(\alpha , \beta) f(\alpha) f(\beta) \, \dd {\text{Re}}(\alpha) \dd {\text{Im}}(\alpha) \dd {\text{Re}}(\beta) \dd {\text{Im}}(\beta),
\end{equation}
where $\hat{a}_A, \hat{a}^\dagger_A$ and $\hat{a}_B, \hat{a}_B^\dagger$ represent the annihilation and the creation operators of Alice's and Bob's systems, respectively. $f(\alpha)$ is obtained from $\hat{f} (\hat{a}_A , \hat{a}_A^\dagger)$ by replacing $\hat{a}_A$ by $\alpha$ and $\hat{a}_A^\dagger$ by $\alpha^*$, and $f(\beta)$ is also obtained in the analogous way. Since the quantum state between Alice and Bob under the condition that the successful events occur can be expressed as a weighted difference of two Gaussian states, it is possible to analytically calculate the expectation values.

\section{C. ADDITIONAL NOISE ANALYSES}\label{sec:additional}
\begin{figure}[tbp]
 \centering
 \includegraphics[keepaspectratio, scale=0.4]{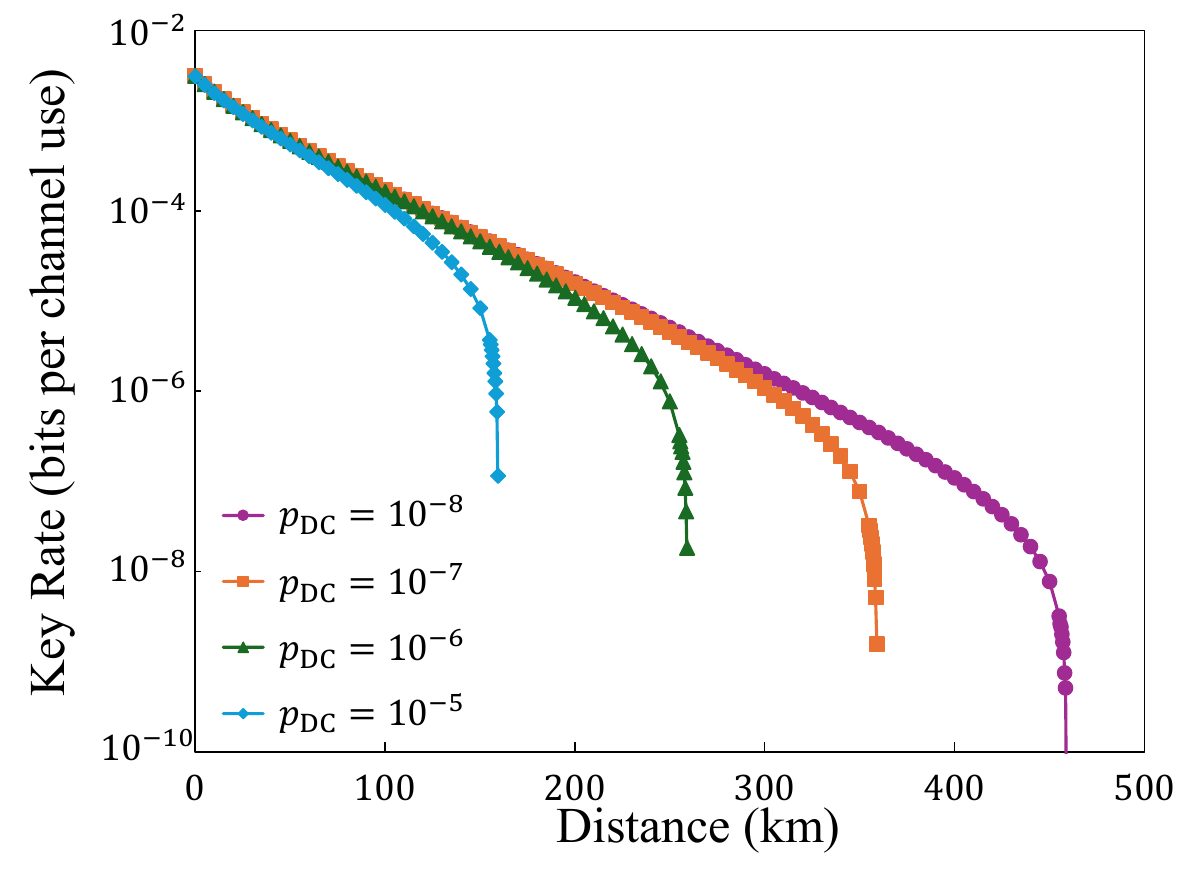}
 \caption{Key rate versus distance $L$ for different dark count probability $p_\text{DC}$. We take $\eta_e^c = 0.7$ and $\beta=0.95$.}
 \label{fig:DifferentPDC}
\end{figure}

\begin{figure}[tbp]
 \centering
 \includegraphics[keepaspectratio, scale=0.4]{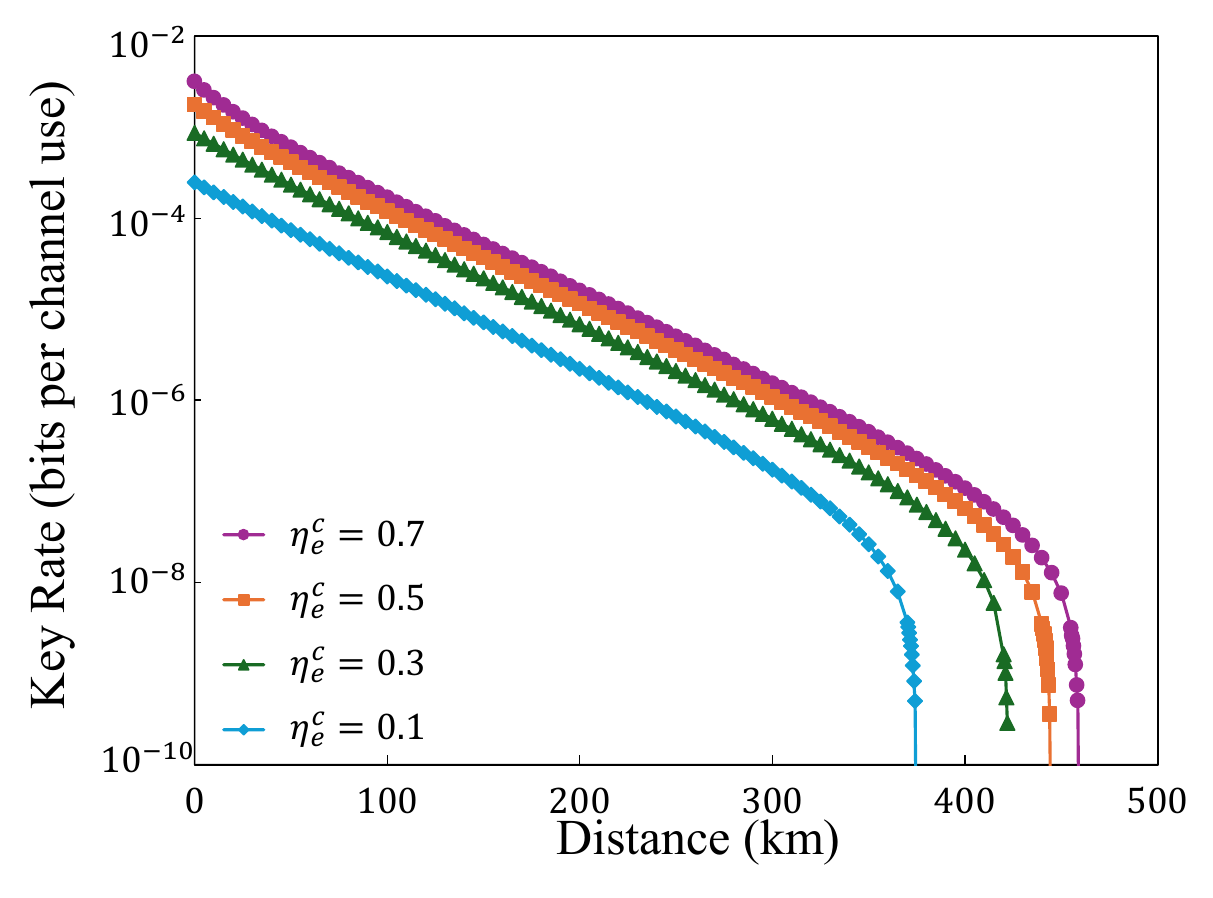}
 \caption{Key rate versus distance $L$ for different efficiency of the on-off detectors $\eta_e^c$. We take $p_\text{DC} = 10^{-8}$ and $\beta=0.95$.}
 \label{fig:DifferentEtaec}
\end{figure}

In Fig.~\ref{fig:DifferentPDC} and Fig.~\ref{fig:DifferentEtaec}, we plot key rates versus the distance $L$ for different dark count probability $p_\text{DC}$ and efficiency of the on-off detectors $\eta_e^c$, respectively. We observe that the dark count probability has a stronger effect on the key rates than the efficiency of the on-off detectors. This is because we use single-photon interference at the central node. When the dark-count probability is comparable to the probability that Alice's or Bob's single photon arrives at the central station, we cannot distinguish whether the successful events are caused by the dark count or the single photon. On the other hand, decrease of efficiency of the on-off detectors corresponds to increase of channel losses. Therefore, the key rate is not changed significantly by a decrease in detection efficiency.

\end{document}